\documentclass[aps,pre,amsmath,amssymb,twocolumn,showpacs]{revtex4}     
\usepackage{graphicx}

\begin{document}
\title{Mixture-like behavior near a liquid-liquid phase transition in simulations of supercooled water}

\author{Megan J. Cuthbertson}
\affiliation{Department of Physics, St. Francis Xavier University,
Antigonish, NS, B2G 2W5, Canada}

\author{Peter H. Poole}
\affiliation{Department of Physics, St. Francis Xavier University,
Antigonish, NS, B2G 2W5, Canada}

\begin{abstract}
In simulations of a water-like model (ST2) that exhibits a liquid-liquid phase transition, we test for the occurrence of a thermodynamic region in which the liquid can be modelled as a two-component mixture.  We assign each molecule to one of two species based on the distance to its fifth-nearest neighbor, and evaluate the concentration of each species over a wide range of temperature and density.  Our concentration data compare well with mixture-model predictions in a region between the liquid-liquid critical temperature and the temperature of maximum density.  Fits of the model to the data in this region yield accurate estimates for the location of the critical point.   We also show that the liquid outside the region of density anomalies is poorly modelled as a simple mixture.
\end{abstract}


\date{\today;   Phys. Rev. Lett., in press}
\maketitle

The possibility that a liquid-liquid phase transition (LLPT) occurs in supercooled water and other tetrahedral liquids (e.g. silicon) continues to be a subject of investigation and debate~\cite{PSES,stanley}.  In the LLPT proposed for water, two phases, a low-density liquid (LDL) and high-density liquid (HDL), become distinct below a critical temperature $T_c$ located in the supercooled regime.  While a LLPT for water has yet to be confirmed experimentally, simulation studies have identified unambiguous LLPTs in several model tetrahedral liquids, including water~\cite{deben}, silicon~\cite{sastry}, and tetrahedrally coordinated colloids~\cite{francis}.

Long before any discussion of LLPTs, there were recurring proposals that the thermodynamic anomalies of water, such as the density maximum, could be understood if the liquid is modelled as a mixture of two ``species" differing in local molecular structure:  one of lower density and disorder, and the other of higher density and disorder~\cite{rontgen}.   Following the emergence of evidence for the continuum nature of the local structure and bonding in the liquid under ambient conditions, mixture models of water faded from prominence~\cite{malenkov}.  However, the proposal of a LLPT has renewed interest in mixture models, in which spontaneous LDL-like and HDL-like structural fluctuations play the role of the mixed species~\cite{connie,pony1}.  Recent experiments have been interpreted as evidence for such mixture-like fluctuations in water~\cite{huang}, although this interpretation is disputed~\cite{clark}.

One commonly discussed mixture model for water~\cite{connie,pony1} is based on the Gibbs free energy $G$ of a binary regular solution, given by,
\begin{eqnarray}
\label{mix}
G&=&G_A(1-X) + G_BX + wX(1-X) \\
&&+ RT[X\ln X + (1-X) \ln (1-X)], \nonumber
\end{eqnarray}
where $T$ is the temperature, $X$ is the concentration of component B, $G_A$ and $G_B$ are the respective free energies of the pure A and pure B liquids, and $R$ is the gas constant.  The energy of mixing is quantified by $w$.  It is further assumed that the two species can interconvert ($A \rightleftharpoons B$), and that the free energy difference between the two pure phases is given by $G_B-G_A=\Delta E -T\Delta S + P\Delta V$.  Here $\Delta E$, $\Delta S$ and $\Delta V$ are respectively the difference in energy, entropy and molar volume of the two pure phases (e.g. $\Delta V=V_B-V_A$), and for simplicity are assumed to be constant with respect to $T$ and pressure $P$.  In this modified regular solution (MRS) model, the equilibrium value of $X=x$ at fixed $P$ and $T$ is determined by minimizing Eq.~\ref{mix} with respect to $X$.  For $w>0$, a liquid-liquid critical point occurs at $T_c=w/2R$, $P_c=(T_c\Delta S-\Delta E)/\Delta V$, and $x_c=1/2$.  This MRS model was originally developed by Rapaport to account for melting line maxima in pure liquids~\cite{rap}, and is a member of a large group of two-state models that have been applied to thermodynamic and relaxation phenomena in one-component liquids~\cite{tanaka, angell} and crystals~\cite{kittel,aptekar}.  

MRS models have illustrated how a LLPT and the thermodynamic anomalies of water may be interrelated.  These models have also been reported to be in quantitative agreement with experimental data for water~\cite{connie,pony1,tanaka}, but such comparisons are complicated by two significant challenges.  First, the four model parameters $w$, $\Delta E$, $\Delta S$, and $\Delta V$ are not known for real water, and so they must be estimated indirectly, e.g. from the properties of the amorphous ices.  Second, in order to compute thermodynamic properties from Eq.~\ref{mix}, the free energy function of a reference state [e.g. $G_A(P,T)$] is required.  By themselves, the four model parameters determine only the ``anomalous" contribution to thermodynamic properties arising from the variation of $x$.  Hence, to fit the model to experimental data without knowledge of $G_A(P,T)$, a ``normal" contribution must first be estimated and subtracted from the data.  This process introduces additional and difficult-to-estimate parameters.  Consequently, the regime of validity of MRS models for describing behavior near a LLPT, if such a regime exists at all, remains uncertain.

In this Letter, we use simulation data to test if a thermodynamic regime exists in which a MRS model can describe a water-like liquid near a LLPT.  We study the ST2 model of water~\cite{st2}, as it provides a context in which the two difficulties described above can be avoided.  First, ST2 exhibits a well-characterized LLPT~\cite{denmin,deben}.  We can therefore determine the four model parameters of the MRS model directly.  Second, we identify a property of the local structure in ST2 that allows us to estimate the concentrations of LDL-like and HDL-like species.  We compare these concentrations directly to the predictions of the MRS model, thus avoiding the need to decompose thermodynamic properties into normal and anomalous contributions. 

\begin{figure}
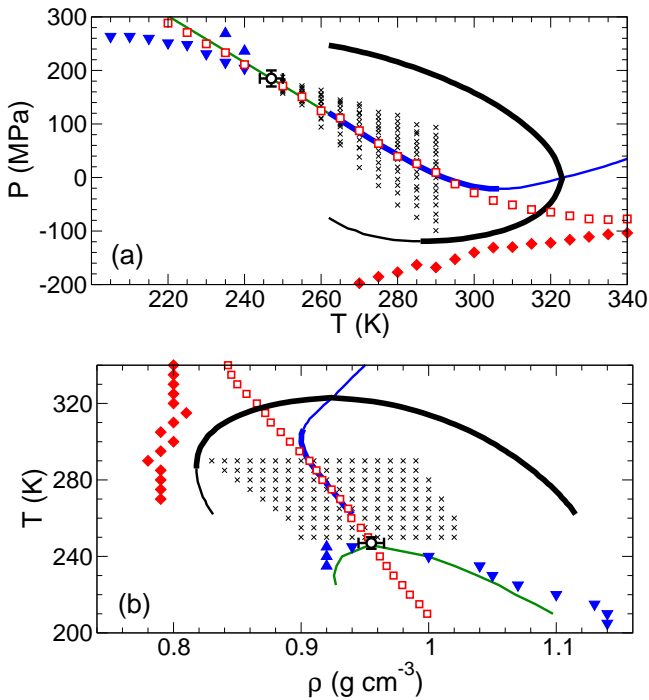
\bigskip
\centerline{\includegraphics[scale=0.33]{fig-1a.eps}}
\medskip
\centerline{\includegraphics[scale=0.33]{fig-1b.eps}}
\caption{(a) $T$-$P$ and (b) $\rho$-$T$ projections of the properties of ST2 water.  
From Ref.~\cite{denmin} we show the line of density maxima (thick black line); density minima (thin black line); $K_T$ maxima (thick blue line); $K_T$ minima (thin blue line); the liquid spinodal (diamonds); the HDL spinodal (down triangles); and the LDL spinodal (up triangles).
We also show the locus along which $x=0.5$ (open red squares); the location of the critical point (open black circle); and the state points falling inside the region $\cal R$ defined in the text (crosses).  
The green line in (a) is a line of slope $m$, and in (b) is the coexistence curve predicted by the MRS model.
}\label{pd}\end{figure}

\begin{figure}\bigskip
\centerline{\includegraphics[scale=0.35]{fig-2a.eps}}
\centerline{\includegraphics[scale=0.35]{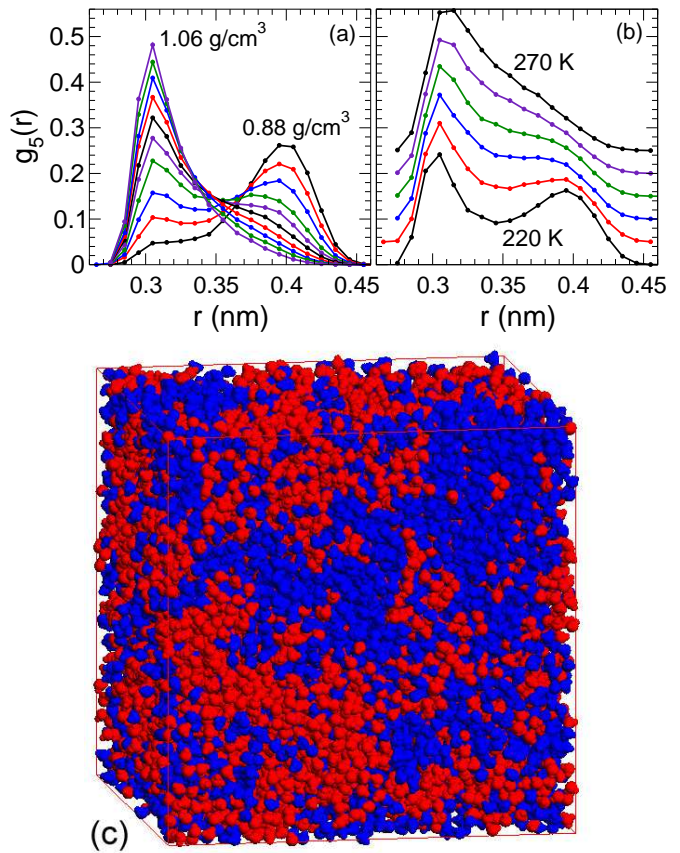}}
\caption{(a) $g_5(r)$ at $T=245$~K, for $\rho=0.88$ to $1.06$~g/cm$^3$ in steps of $0.02$~g/cm$^3$.  (b) $g_5(r)$ at $\rho=0.96$~g/cm$^3$, for $T=220$ to $270$~K in steps of $10$~K.  For clarity, curves for $T> 220$~K are successively shifted upward by 0.05.  (c) A snapshot of a system of $N=13\,824$ ST2 molecules at $T=245$~K and $\rho=0.96$~g/cm$^3$;  at this state we find $P=191$~MPa.  Blue molecules have $r_5>0.35$~nm; red molecules have $r_5<0.35$~nm.}\label{g5}\end{figure}

Our results are based on molecular dynamics simulations of a system of $N=1728$ ST2 molecules.  Our runs are conducted at fixed volume $V$, and $T$ is controlled using Berendsen's method~\cite{beren}.  Long-range contributions to electrostatic interactions are approximated using the reaction field method.  We study a wide range of states:  from $T=220$ to 400~K, in 5~K steps; and from density $\rho=0.8$ to $1.1$~g/cm$^3$, in steps of $0.01$~g/cm$^3$.  Complete details of our simulation procedure are as described in Ref.~\cite{denmin}.  Fig.~\ref{pd} summarizes the known phase behavior of ST2.  As reported in Ref.~\cite{denmin}, a liquid-liquid critical point occurs in the vicinity of $T_c=245$~K, $P_c=185$~MPa and $\rho_c=0.94$~g/cm$^3$.  

First, we seek a criterion for assigning molecules to LDL-like and HDL-like species.  In Fig.~\ref{g5}, we analyze the liquid structure near the critical point in terms of the distance $r_5$ from the O atom of each molecule to its fifth-nearest neighbor.  Following Ref.~\cite{saika}, we define $g_5(r)$ such that $\rho g_5(r)$ is the average density of fifth-nearest neighbors of an O atom at the origin, as found in a volume element at a distance $r$.  So defined, the conventional pair correlation function $g(r)=\sum_{i=1}^{\infty}g_n(r)$.  In the range of $\rho$ studied here, fifth-nearest neighbors are located over a range of distances that span the first minimum in the O-O pair correlation function, and thus $r_5$ is an indicator of the degree to which the tetrahedral structure of the first coordination shell is disrupted by additional neighbors.  At $T=T_c$ we find that $r_5$ is typically greater than 0.35~nm in the LDL phase ($\rho<\rho_c$), while in the HDL phase ($\rho>\rho_c$) $r_5$ is typically less than 0.35~nm [Fig.~\ref{g5}(a)].  Near $\rho=\rho_c$ for $T<T_c$, $g_5(r)$ has a bimodal shape indicating the presence of distinct populations of LDL-like and HDL-like coordination environments [Fig.~\ref{g5}(b)].  We therefore adopt $r_5$ as a local order parameter for assigning molecules to two species:  ``A" molecules are LDL-like and have $r_5>0.35$~nm;  ``B" molecules are HDL-like and have $r_5<0.35$~nm.  Fig.~\ref{g5}(c) shows an equilibrium configuration from a separate simulation of $N=13\,824$ ST2 molecules at a state within error of the critical temperature and density.  This image, in which A and B molecules are shown in different colors, confirms the presence of large, spatially correlated clusters of each species, as expected near a critical point.  

\begin{figure}
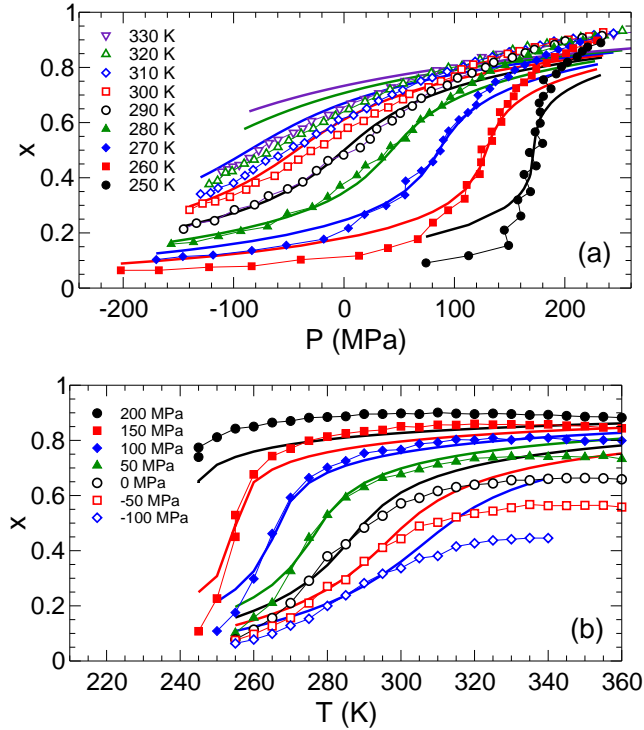
\bigskip
\centerline{\includegraphics[scale=0.34]{fig-3a.eps}}
\medskip
\centerline{\includegraphics[scale=0.34]{fig-3b.eps}}
\caption{(a) Isotherms of $x$ as a function of $P$.  
(b) Isobars of $x$ as a function of $T$. 
Solid curves are the corresponding predictions of Eq.~\ref{mix}.  
}\label{x}\end{figure}

We evaluate $x$, the equilibrium number concentration of B molecules in the system, at each $(\rho,T)$ state point as a time average over the equilibrium configurations generated during the run.  We also evaluate $P$, to allow us to analyze $x(P,T)$ as well as $x(\rho,T)$.  Our results for $x$ are shown as isotherms as a function of $P$ in Fig.~\ref{x}(a), and as isobars as a function of $T$ in Fig.~\ref{x}(b).  Approaching the critical point, the qualitative behavior of $x$ is as expected for a liquid mixture approaching a LLPT:  Both isotherms (as $P\to P_c$) and isobars (as $T\to T_c$) become more steeply sloped.

\begin{figure}\bigskip
\centerline{\includegraphics[scale=0.34]{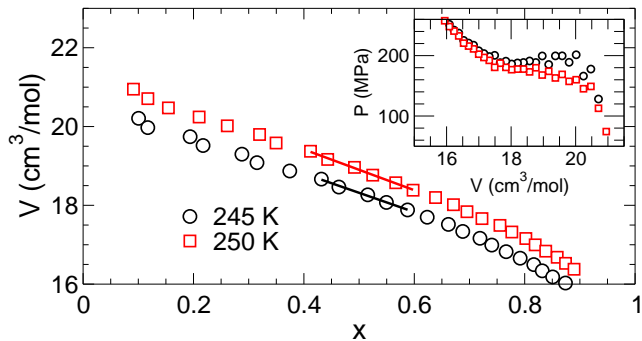}}
\caption{Isotherms near $T_c$ of $V$ as a function of $x$.  The lines are linear fits to the data for $0.4>x>0.6$.  For clarity, the data for $T=245$~K have been shifted downward by 0.5~cm$^3$/mol.  Inset:  Isotherms of $P$ versus $V$ straddling $T_c$.
}\label{Vx}\end{figure}

Next we evaluate the parameters of the MRS model appropriate for ST2.  We confirm the estimates of Ref.~\cite{denmin} for $T_c$ and $P_c$ by examining two $P$-$V$ isotherms straddling $T_c$ (inset, Fig.~\ref{Vx}).  Along the $T=250$~K isotherm, $P$ is monotonic in $V$, while the first sign of a ``van der Waals loop" occurs along the $T=245$~K isotherm.  Given the scatter of the data along these isotherms, we estimate $T_c=247\pm 3$~K and $P_c=185\pm 15$~MPa.


We estimate $\Delta V$ by noting that in the MRS model $V=V_n+\Delta V x$, where $V_n$ is the ``normal" or non-singular contribution to $V$,  and $\Delta V x$ is the ``anomalous'' contribution due to the variation of $x$~\cite{pony1}.  $V_n$ in general depends on both $P$ and $T$.  However,  along the critical isotherm near $V_c=1/\rho_c$, both $P$ and $T$ are constant (inset, Fig.~\ref{Vx}), and therefore $V_n$ is constant.  Accordingly, we estimate $\Delta V$ from the slope of $V$ versus $x$ along the critical isotherm in the interval $0.4<x<0.6$.  As we will see below, this range of $x$ spans the states near $\rho_c$.  By averaging the slopes (and their errors) obtained for $T=245$ and $250$~K, we estimate $\Delta V=-5.0\pm 0.2$~cm$^3$/mol (Fig.~\ref{Vx}).  

As $T\to T_c$ from above, the MRS model predicts that the value of the isothermal compressibility $K_T$ is increasingly dominated by a term containing $(\partial x/\partial P)_T$, which diverges at the critical point~\cite{pony1}.  In the model, the maxima of isotherms of $(\partial x/\partial P)_T$ occur at $x=1/2$ for all $T$.  As a consequence, the locus along which $K_T$ is a maximum, which converges to the ``Widom line" as $T\to T_c$~\cite{widom}, should also converge with the $x=1/2$ locus in the region approaching the critical point.  The set of points satisfying $x=1/2$ is plotted in Figs.~\ref{pd}(a) and (b), along with the locus of $K_T$ maxima reported in Ref.~\cite{denmin}.  We find that the $x=1/2$ locus is in excellent agreement with the locus of $K_T$ maxima for $T<290$~K, and also that the location of the critical point is consistent with the model prediction of $x_c=1/2$ in both the $P$-$T$ and $\rho$-$T$ planes.  Computing the value of $\rho$ at which $x=1/2$ at $T=T_c$ predicts $\rho_c=0.955\pm 0.01$~g/cm$^3$.

We also find that the $x=1/2$ locus, both above and below $T_c$, closely approximates a straight line in the $P$-$T$ plane for $T<290$~K.  In the MRS model, both the $x=1/2$ locus for $T>T_c$, as well as the coexistence curve for $T<T_c$,  follow a straight line in the $P$-$T$ plane given by $P^\ast=(\Delta S/\Delta V)T-(\Delta E/\Delta V)$~\cite{pony1}.  From a linear fit to the $x=1/2$ locus from 250 to 280~K, we obtain the slope $m=\Delta S/\Delta V=-4.3\pm 0.2$~MPa/K.  We note that the resulting prediction for the coexistence line (with a Clapeyron slope of $m$) as expected lies between the LDL and HDL spinodal lines in Fig.~\ref{pd}(a).

Our estimates for $\Delta V$, $T_c$, $P_c$, and $m$ completely determine the four model parameters ($\Delta V$, $w$, $\Delta E$, and $\Delta S$) required for Eq.~\ref{mix}.  We use these values to obtain the MRS model prediction for $x(P,T)$, and compare the results to the data for $x$ from our simulations (Fig.~\ref{x}).  We find that inside a region $\cal R$, defined approximately as $0.25<x<0.75$ and $250\,{\rm K}<T<290\,{\rm K}$, the MRS model is in good agreement with the values of $x$ computed from our simulations.  At the boundaries of $\cal R$ and beyond, the agreement rapidly degrades.

To test the robustness of the agreement between the MRS model and our data in the region $\cal R$, we carry out a least-squares fit of the model to our data for $x$ in this region, allowing all four model parameters to vary.  We select all distinct pairs of isotherms of $x$ in $\cal R$ that are at least 15~K apart, and fit the model to each of the 21 data subsets so defined.  This gives 21 separate estimates for each fit parameter, from which we compute the mean and the standard deviation. The results are in excellent agreement with the values obtained above:
$T_c=247\pm 3$~K,
$P_c=181\pm 11$~MPa,
$\Delta V=-5.2 \pm 0.6$~cm$^3$/mol, and 
$m=-4.3 \pm 0.2$~MPa/K.

Our results thus demonstrate that a mixture model can indeed provide a quantitatively accurate description of a water-like liquid, in the specific case that the liquid exhibits a LLPT.  Our MRS model successfully predicts the concentrations of LDL-like and HDL-like structural fluctuations in the region $\cal R$, which lies inside the locus of density extrema, above $T_c$, and spans the range $x=0.5\pm0.25$ centered on the Widom line (Fig.~\ref{pd}).  The signatures for the onset of this ``mixture-model regime" are the merging of the Widom line with the $x=1/2$ locus, and the observation of a linear Widom line in the $P$-$T$ plane.  These signatures may be useful for assessing other water-like liquids (either in simulations or experiments) for mixture-like behavior.  

Outside $\cal R$, the MRS model does a poor job estimating $x$.  
Fig.~\ref{pd}(a) shows that the high-$T$ boundary of $\cal R$ is in the vicinity of $290$~K, which is $90\%$ of the highest $T$ ($323$~K) reached by the line of density maxima for ST2 water.  For real water, $90\%$ of the temperature of maximum density at ambient pressure (277~K) gives an estimate of $249$~K ($-24$~C) for the highest $T$ at which mixture-like behavior might be observed experimentally.
This estimate supports the view that mixture models are not appropriate for interpreting the behavior of real water at ambient $T\simeq 300$~K~\cite{clark}.  

We also note that the MRS model fails for $T<T_c$.  Fig.~\ref{pd}(b) shows that the coexistence curve predicted by the MRS model~\cite{pony1} is too narrow since, in violation of thermodynamics, it lies inside the estimate of the LDL and HDL spinodal lines.  The MRS model is a mean-field theory, and thus the shape of the coexistence curve near $T_c$ obeys $(\rho-\rho_c)\propto [(T_c-T)/T_c]^\beta$, with $\beta=1/2$.  However, the LLPT in ST2 water has been shown to belong to the 3D Ising universality class, for which $\beta\simeq 0.327$~\cite{deben}.  Hence it is to be expected that the 
MRS model will underestimate the density difference between the coexisting phases as $T$ decreases below $T_c$.  In Fig.~\ref{x} we also note that there are significant deviations between the model and the data in the limits of large and small $x$ for $T>T_c$, highlighting that the nearly pure A and B phases are poorly described by the model.  This may also contribute to the discrepancy between the model and the data for the coexistence curve for $T<T_c$.

However, within the mixture-model regime defined by $\cal R$, our work shows that the examination of a local structural property  as a function of $T$ and $P$ can yield accurate information concerning the location of the Widom line and the critical point of a LLPT.  In our case, the local structure is quantified in terms of $x$, which is determined by the $r_5$ values of individual molecules.  A number of simulation studies have used the behavior of $g_5(r)$ and related measures as evidence for a LLPT, in both tetrahedral~\cite{sastry,saika} and non-tetrahedral liquids~\cite{bonev}.  Our results validate this approach, and further, confirm that the structures relevant to the LLPT in water-like liquids are highly localized, extending no farther than the second coordination shell.

If a mixture-model regime exists for real water, our results suggest that it will be found 
in the vicinity of the Widom line.  States on 
the Widom line have yet to be studied in experiments on bulk supercooled water, due to the onset of rapid ice crystallization.  However, for tetrahedral liquids in which the region of the Widom line is accessible, which may be the case for nanoconfined water~\cite{liu}, our results demonstrate that probes of local molecular structure, in concert with mixture-model concepts, can be used to elucidate the properties of a LLPT and its associated critical point.

We thank K. Fraser and C. Creelman for useful discussions; ACEnet for computing resources; and NSERC, AIF and the CRC program for financial support.

\end{document}